\documentclass[aps,floatfix,showpacs,epsfig]{revtex4}
\usepackage{pdfpages}
\usepackage{epsf}
\usepackage{subfigure}
\usepackage{amsmath}
\usepackage{epstopdf}
\usepackage{mathrsfs}
\usepackage{natbib}
\usepackage{amssymb,latexsym}
\usepackage{dcolumn}
\usepackage{graphicx}

\def\be{\begin{equation}}
\def\ee{\end{equation}}
\def\ba{\begin{eqnarray}}
\def\ea{\end{eqnarray}}

\graphicspath{{images/}}
\renewcommand{\L}{Lema\^{i}tre}
\begin{document}
\title{\large \bf  The spherical perfect fluid collapse with pressure in the cosmological background}
\author{Rahim Moradi}
\affiliation{Department of Physics, Shahid Beheshti University, G.C., Evin, Tehran 19839, Iran }
\email{Ra_Moradi@sbu.ac.ir}

 \author{Javad T. Firouzjaee}
\affiliation{ School of Physics and School of Astronomy, Institute for Research in Fundamental Sciences (IPM), Tehran, Iran }
 \email{j.taghizadeh.f@ipm.ir}
\author{Reza Mansouri}
\affiliation{Department of Physics, Sharif University of Technology,
Tehran, Iran and \\
  School of Astronomy, Institute for Research in Fundamental Sciences (IPM), Tehran, Iran}
 \email{mansouri@ipm.ir}

\begin{abstract}
We have constructed a spherically symmetric structure model in a cosmological background filled with perfect fluid with
non-vanishing pressure and studied its quasi-local characteristics. This is done by using the \L\ solution of the Einstein
equations and suggesting an algorithm to integrate it numerically. The result shows intriguing effects of the pressure inside
the structure. The evolution of the central black hole within the FRW universe, its
decoupling from the expanding parts of the model, the structure of its space-like apparent horizon, the limiting case of
the dynamical horizon tending to a slowly evolving horizon, and the decreasing mass in-fall to the black hole is also studied. The
quasi-local features of this cosmological black hole may not be inferred from the weak field approximation although the gravity
outside the structure is very weak.
\end{abstract}
\maketitle
\section{introduction}

The term cosmological black hole is used to describe a collapsing structure within an otherwise expanding universe. Since the early beginning
of the discovery of the expansion of the universe people have been looking for models describing an overdense region in a cosmological
background (\cite{McVittie}, see also \cite{cosmological black hole}). The initial expansion of such an overdense region will finally decouple
from the background and collapses to a dynamical black hole. The resulting structure and its central black hole, despite the very weak gravitational
field outside it, differ from the familiar Schwarzschild one in being neither static no asymptotically flat \cite{man}. Therefore, such cosmological
structures, if based on exact solutions of general relativity and not produced by a cut-and-paste technology, are very interesting laboratories to
study not only general relativistic structures, their quasi-local features such as mass and horizons \cite{man}, but the validity of the weak field
approximation in the presence of the very weak gravitational fields relevant to non-local concepts within general relativity \cite{mojahed}. After
all, the universe is evolving and asymptotically not Minkowskian. Therefore, one needs to have a dynamical model for a black hole to be
compared with the familiar results in the literature on black holes within a static and asymptotically flat space-time \cite{waldbook} where
global concepts such as event horizon are not defined.  The need for a local definition of black holes and their horizons has led us to concepts
such as Hayward's trapping horizon {\cite{Hayward94}, isolated horizon \cite{ashtekar99}, Ashtekar and Krishnan's dynamical
horizon (DH) \cite{ashtekar02}, and Booth and Fairhurst's slowly evolving horizon {\cite{booth04}. \\

Now, a widely used metric to describe the gravitational collapse of a spherically symmetric dust cloud is the so-called
Tolman-Bondi-\L (LTB) metric \cite{LTB}. It was pointed out in \cite{man, mangeneral} that the model admits cosmological
black holes. These models may be extended to a perfect fluid with a non zero pressure, the so-called \L \ models \cite{hellaby}. Our
interest is now using these cosmological solutions to construct dynamical black holes within a FRW expanding universe, and study
their characteristics. To do this, we have to avoid any cut-and-paste method of finding the solution. Section II is an introduction to these inhomogeneous
prefect fluid cosmological models and how numerically integrate the field equations leading to a dynamical structure within an
otherwise FRW universe. In section III the result of the numerical integration is reported expressing the main characteristics of our
model assuming different pressure profiles. We will then discuss the result in section IV. Throughout the paper we assume $8\pi G = c = 1$.

\section{General spherically symmetric solution}

   Consider a general inhomogeneous spherically symmetric spacetime \cite{hellaby} filled with a perfect fluid and a metric expressed in the comoving coordinates, $x^{\mu} = (t,\,r,\,\theta,\,\phi)$:
\begin{equation}\label{metric}
ds^2  = -e^{2\sigma} \, dt^2  + e^{2\lambda} \, dr^2 + R^2 \, d\Omega^2 \;,
\end{equation}
 where $\sigma = \sigma(t,r)$, $\lambda = \lambda(t,r)$ are functions to be determined, $R = R(t,r)$ is the physical radius, and $d\Omega^2 = d\theta^2 + \sin^2 \theta \, d\phi^2$ is the metric of the unit 2-sphere.  The energy momentum tensor of the perfect fluid is given by
 \begin{equation}\label{mT}
   T^{\mu\nu} = (\rho + p) \, u^{\mu} \, u^{\nu} + g^{\mu\nu}p \;,
 \end{equation}
 where $\rho = \rho(t,r)$ is the mass-energy density, $p = p(t,r)$ is the  pressure,  and $u^{\mu} = (e^{-\sigma}, 0, 0, 0)$ is the perfect fluid four-velocity.

\subsection{Field Equations} \label{FEq}

  In addition to the Einstein field equations, $G^{\mu\nu} = \kappa \,T^{\mu\nu} - g^{\mu\nu}\Lambda$, we will use the
   conservation equations in the form \newline
    \begin{equation}
   \frac{2 e^{2\sigma}}{(\rho + p)} \, \nabla_\mu T^{t\mu}  = \dot{\lambda}
      + \frac{2 \dot{\rho}}{(\rho + p)} + \frac{4 \dot{R}}{R} \; = 0
      \label{gtt}
       \end{equation}
   \begin{equation}
   \frac{e^{\lambda}}{(\rho + p)} \, \nabla_\mu T^{r\mu}  = \sigma' + \frac{p'}{p + \rho} = 0,
      \label{grr}
 \end{equation}
where the dot means the derivative with respect to $t$, and the prime means the derivative with respect to $r$.
The Einstein equations lead finally to following equations
\begin{equation}\label{ev}
   \frac{\partial}{\partial r} \left[ R + R \dot{R}^2 e^{-2\sigma} - R  R'^2 e^{-\lambda} - \frac{1}{3} \Lambda R^3 \right] = \kappa \rho R^2 R', \;
 \end{equation}
  and
\begin{equation}\label{Pr}
   \frac{\partial}{\partial t}\left[ R + R \dot{R}^2 e^{-2\sigma}-RR'^2 e^{-\lambda}-\frac{1}{3}\Lambda R^3\right] = -\kappa p R^2\dot{ R}. \;
   \end{equation}
 The term in the brackets is related to the Misner-Sharp mass, $M,$ defined by
 \begin{equation}
   \frac{2M}{R} = \dot{R}^2 e^{-2\sigma} - R'^2 e^{-\lambda} + 1 - \frac{1}{3} \Lambda R^2 \;.
   \label{mR}
 \end{equation}
 Eqs (\ref{ev}) and (\ref{Pr}) may now be written as
 \begin{equation}
   \kappa \rho  = \frac{2M'}{R^2 R'} ~,
   \kappa  p  = -\frac{2\dot{M}}{R^2 \dot{R}} ~.   \label{pressure}
 \end{equation}
 We may write Eq (\ref{mR}) in the form of an evolution equation of the model:
 \begin{equation}
   \label{Ev}
      \dot{R} = \pm e^{\sigma} \sqrt{\frac{2M}{R} + f + \frac{\Lambda R^2}{3}} ~, \\
 \end{equation}
where
\begin{equation}
   \label{f}
      f(t, r) = R'^2 e^{-\lambda} - 1~
 \end{equation}
 is the curvature term, or twice the total energy of test particle at $r$ (analogous to $f(r)$ in the LTB model). Note that $R(t,r)$ can not be directly obtained from this
 equation because of the unknown functions $\lambda$, $\sigma$, and $M$.

   The metric functions $g_{tt}$ and $g_{rr}$ may be obtained by integrating (\ref{gtt}) and (\ref{grr}):
 \begin{equation}
   \label{si}
      \sigma  = \ c(t) - \int_{r_0}^r  \frac{ p' \, \ dr}{(\rho + p)}|_{t=const}
      = \sigma_0 -\int_{\rho_0}^\rho \frac{(\frac{\partial p}{\partial\rho})}{(\rho + p(\rho))} \, d\rho~|_{t=const},
 \end{equation}
 and
 \begin{equation}
   \label{lamb}
      \lambda  = \lambda_0(r) - 2 \int_{\rho_0}^\rho
      \frac{d\rho}{(\rho + p(\rho))}
      - 4 \ln \left( \frac{R}{R_0} \right)~|_{t=const},
 \end{equation}
 where $c(t)$ and $\lambda_{0}(r)$ are arbitrary functions of integration (see \cite{hellaby} for more details). In the case of $c(t)$ it is easily seen that requiring our coordinates to lead
 to the LTB synchronous ones for $p = 0$ leads to $c(t)=0$. We notice also that according to (\ref{f}) and the LTB coordinate conditions, the choice of
 $\lambda_{0}(r)$ is equivalent to the choice of $f(t_0,r)=f_0(r)$. One may  prefer to choose $f_0(r)$ and then calculate $\lambda_0(r)$
 from $e^{\lambda_0} = R_0'^2/(1 + f_0)$.\\
  We have therefore 5 unknowns $p, \rho, \sigma, \lambda$, and $R$, four dynamical equations $\dot{\rho}$, $\dot{M}$, $\dot{\lambda}$,
and $\dot{R},$ in addition to an equation of state $p = p(\rho)$, and the definition of the mass $M$ (\ref{mR}).
This defines a numerical algorithm to find solutions for the dynamics of our spherical structure after assuming the initial conditions.

 \subsection{Construction of the \L\ Model}
 \label{alg}

   To generate a \L\ model in the general case we need a numerical procedure. We first specify the arbitrary initial functions, $R_0(r)=R(t_0,r)$,
    $\rho_0(r)= \rho(t_0,r)$, $\lambda_0(r)= \lambda(t_0,r)$ , $\sigma_0(r)= \sigma(t_0,r)$, and the equation of state $p(\rho)$. We
 then integrate the dynamical equations for constant $t$ or $r$ in the following order:

 \begin{enumerate}
 \item   Choose an initial time  $t_0$, and specify $R(t_0, r) = R_0(r)$; $R_0'$ is then also known form the derivative of $ R_0(r)$ with respect to $r$ at $t=t_0$;
  \item     Specify $M(t_0, r) = M_0(r) $ at $ t=t_0$;
  \item  Once $M_0(r)$ and $ R_0(r)$ are specified, $\rho_0(r)$  may be determined from (\ref{pressure});
 \item   Select an equation of state, $p = p(\rho)$;
 \item   Choose $\lambda(t_0, r)$, or  choose first $f(t_0,r)=f_0(r)$ and then calculate $\lambda_0(r)$ from $e^{\lambda_0} = R_0'^2/(1 + f_0)$;
  \item   $\sigma(t_0, r)$ can then be obtained by integrating (\ref{si}) along $t = t_0$.
 \end{enumerate}
 We have now specified how to determine all the needed initial functions at the time $t_0$. Therefore, their $r-$derivatives are also known. The
 time evolution of the metric functions along the worldlines of constant $r$ may then be calculated.  This is done in the following way:
 \begin{enumerate}
 \item   From equation (\ref{Ev}) we obtain $\dot{R}$

 \item  Eq (\ref{pressure}) then gives us $\dot{M}$:
 \begin{equation}
   \dot{M} = \frac{- \kappa p \, \dot{R} R^2}{2} ~;
 \end{equation}
 \item  Combining Eq (\ref{gtt}) and $G_{01} $ allows us to eliminate $\sigma'$. Then substituting  $\dot{\lambda}$ from (\ref{grr}), we arrive
  at $\dot{\rho}$:
 \begin{equation}
   \dot{\rho} = - p' \, \frac{\dot{R}}{R'}
      - (\rho + p) \left[ \frac{\dot{R}'}{R'} + \frac{2 \dot{R}}{R} \right] ~;
      \label{rhodot}
 \end{equation}
 \item    From the equation of state $p = p(\rho)$ we obtain $\dot{p}$
 \begin{equation}
   \dot{p} = \frac{dp}{d\rho} \, \dot{\rho}  ~;
 \end{equation}
 \item   Eq (\ref{grr}) may be combined with (\ref{rhodot}) to give
 \begin{equation}
   \dot{\lambda} = \frac{2}{R'} \left( \frac{ p' \dot{R}}{(\rho + p)} + \dot{R}' \right)~;
 \end{equation}
 \item   Using the initial values for $t = t_0$, we are then in a position to solve the above 5 differential equations numerically to obtain $R(t,r)$,
 $M(t,r)$, $\rho(t,r)$, $p(t,r)$ and $\lambda(t, r)$ for every $t$ and $r$. Note that in each step the spatial derivatives $\dot{R}'$, $p'$
 and $M'$ needs to be determined;
 \item   Finally, $\sigma(t,r)$ is obtained from Eq (\ref{si}) by integrating along constant $t$.
 \end{enumerate}
 Notice that we have chosen 4 initial functions, $R_0(r)$, $\rho_0(r)$, $\lambda_0(r)$ and $\sigma_0(t)$, as well as the equation of state $p(\rho)$.

\section{equation of state and the results}

We are now ready to specify the equation of state and integrate the model to see its characteristics. To have a comparative discussion of the
results we consider two types of equation of state: a perfect fluid with a constant state function, $p=w\rho$, and a more general case with
the equation of state $p=w s(r)\rho$ matching our needs for a structure with pressure inside the structure and a pressure-less matter dominated universe far from
the structure. We may then choose the function $ s(r)$ in a way that the pressure becomes zero at infinity, i.e. at  $r>>r_0$. A suitable choice is
$s(r)=e^{-\frac{r}{r_0}}$ where the order of magnitiude $r_0$ is the distance of void form the center (boundary of expanding and collapsing phase). This is a more realistic model to describe a black hole collapse within the FRW universe and to see the effect of the inside
pressure while the universe outside is matter dominated with no pressure.\newline{}

The model we envisage starts from a small inhomogeneity within a FRW universe. The density profile should be such that the metric outside the structure
tends to FRW independent of the time while the central overdensity region undergoes a collapse after some initial expansion. At the initial conditions, where the
density contrast of the overdensity region is still too small, we may assume that the metric is almost FRW or LTB; the density contrast and the
pressure does not play a significant role. The dynamics of \L\ universe will give us anyhow the expected structure at late times. To choose the initial
conditions at the time $t_0$, we will therefore use a LTB solution with a negative curvature function. We have in fact tried both examples of LTB
or FRW initial data and received no significant difference between the final \L\ solutions. \\
Now, let us choose the the two initial functions $f(r)=f(t_0,r)$ and $M(r)=M(t_0,r)$ in the following way to achieve an asymptotically FRW final solution:

\begin{equation}
f(r)=-\frac{1}{b}re^{-r},
\end{equation}
\begin{equation}
M(r)=\frac{1}{a}r^{3/2}(1+r^{3/2}).
\end{equation}
Far from the central overdensity region we have

\begin{equation}
lim_{r\rightarrow\infty}f(r)=0,
\end{equation}
\begin{equation}
lim_{r\rightarrow\infty}M(r)=\frac{r^3}{a},
\end{equation}
showing the asymptotically FRW behavior of the initial conditions. The corresponding LTB solution of Einstein equations now gives us $R(r)=R_0(r)$ at
the initial time $t_0$. Assuming an equation of state is now enough to numerically calculate the necessary dynamical functions of the model.
Specifically, by looking at $\dot{R}(t,r)$ and $\rho(t,r)$ we may extract informations of how the central region starts collapsing after the initial
expansion and how a black hole with distinct apparent and event horizons develops while the outer region expands as a familiar FRW universe. We may also
find out the difference to the case of the pressure-less model. It will also show if and how the very weak gravity outside the collapsed structure
affects the dynamic of the central structure in comparison to the familiar Schwarzschild model. The results of the numerical calculation for both
equation of states are given in the following sections.\\

\subsection{The density behavior}

 The density profiles for both equation of states as a function or $t$ and $r$ are given in Figs.(\ref{den1}) and (\ref{den2}). A comparison of
these figures shows the effect of the pressure on the development of the central black hole. Obviously in case of non-vanishing pressure outside
the structure the collapse is more highlighted with a more steep density profile. The over-density region in the collapsing phase is always
separated from the expanding under-density region through a void not expressible in these figures. We will consider the deepest place of the
void as the boundary of the structure. This boundary is always near by the boundary between the contracting and the expanding region of the
model structure.\newline{}

\begin{figure}[h]
\includegraphics[width = 8cm]{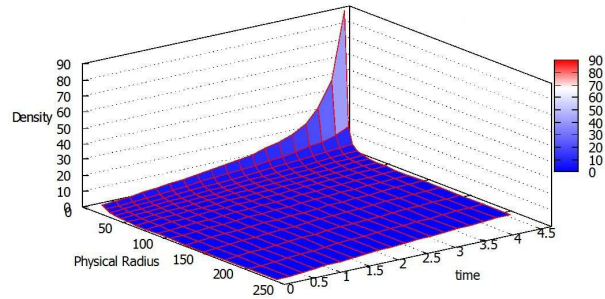}
\caption{ \label{den1} Density evolution of our cosmological black hole for the perfect fluid with the equation of state $p=w\rho$.}
\end{figure}

\begin{figure}[h]
\includegraphics[width = 8cm]{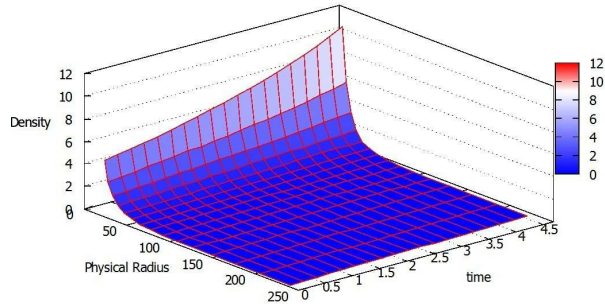}
\caption{ \label{den2} Density evolution of our cosmological black hole for the perfect fluid with the equation of state $p=w\rho s(r)$. Note the less significant central density and the more flat density profile near to the center of the structure.}
\end{figure}

\subsection{ The pressure effect}

Figs.(\ref{2kol}) and (\ref{22kol}) show the behavior of the collapsing and the expanding regions for the equation of state $p = w\rho$ by
depicting the corresponding \L\ Hubble parameter $\dot R/R$ versus the physical radius $R$. Figs. (\ref{1kol}) and (\ref{11kol}) show
similar data for the equation of state $p = ws(r)\rho$. Note that the function $s(r)$, as defined to get matter dominated FRW universe
at far distances, has no significant effect, and the qualitative behavior of the dynamics of the physical radius is independent of it.
Therefore, as far as we are interested in the qualitative features of the model, we will just use the simple equation of state with
$s(r)=1$.\\
The place of separation between the expanding and collapsing region defined by $\dot{R}>0$ and $\dot{R}<0$ is almost coincident with
the place of the void where we have defined as the boundary of the structure. Now, from the  figures we realize that the effect of
the pressure in different regions of the model and its comparison to the homogeneous FRW model is an intriguing one. As we know already
from the Friedman equations in FRW models, the pressure adds up to the density and has an attractive effect slowing down the expansion
leading to a more negative acceleration ($\frac{\ddot{a}}{a}=-\frac{1}{6}(\rho+3 p)$). This is evident from the figures at distances far
from the center where our model tends to an FRW one. Whereas within the structure where we have a contracting overdensity region the
behavior is counter-intuitive. Except for the case of vanishing pressure, in all the other cases the pressure begins somewhere to act
classically like a repulsive force opposing the collapse of the structure. To see this more clearly, we have also depicted the acceleration
in Fig.(\ref{accelaration}). As we approach distances near to the center, the negative acceleration in the FRW limit and even inside the
void, gradually increases to positive values, meaning that somewhere within the structure the contraction of the structure
slows down due to the pressure like a classical fluid. Therefore, the pressure effect begins somewhere within the structure to act like a
repulsive force in contrast to the outer regions where its attractive nature dominates. Note that the central black hole and its horizon has 
a much smaller radius than the region of the repulsive pressure effect we are discussing.

\begin{figure}
  \centering
  \mbox{
    \subfigure[The overall scheme: $\dot{R}>0$ and $\dot{R}<0$ show the expanding and collapsing regions, respectively.\label{2kol}]{\includegraphics[width=0.5\linewidth]{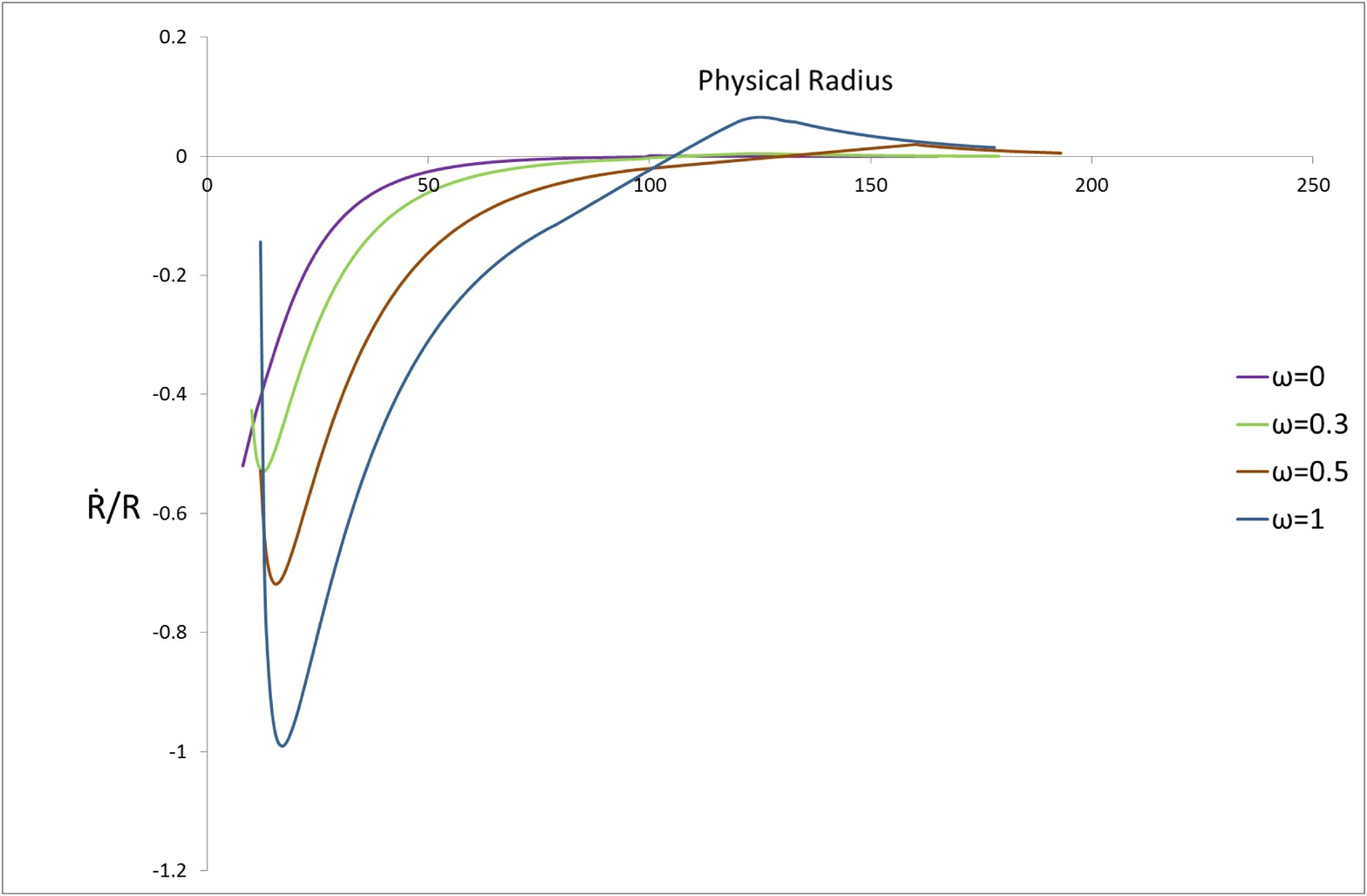}} \quad
   \subfigure[A magnified scheme related to the regions near to the center\label{1kol}]{\includegraphics[width=0.5\linewidth]{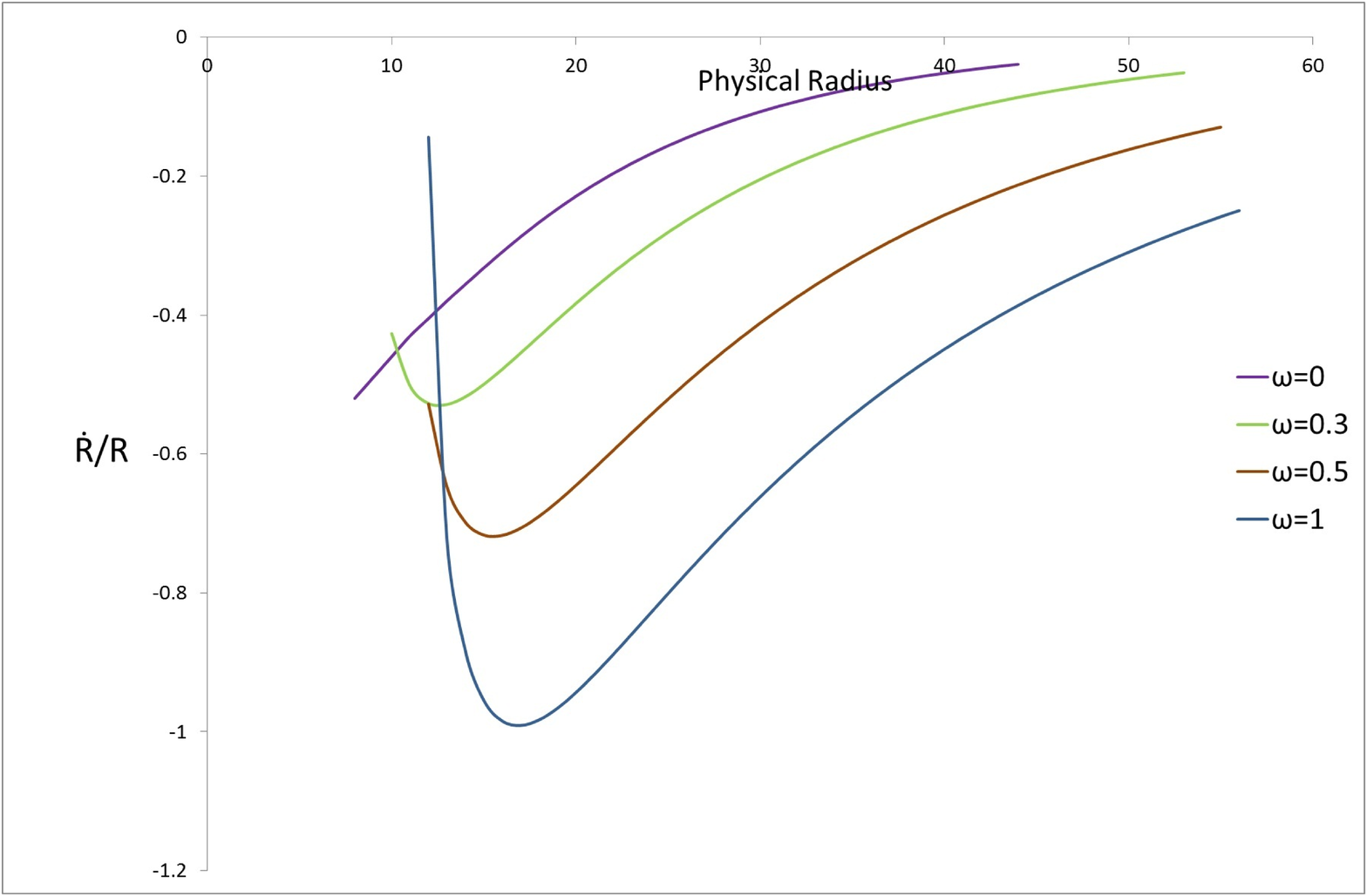}}
  }
  \caption{The behavior of the Hubble parameter $\dot{R}/R$  in the case of $p=w\rho$. Evidently the pressure slows down the
collapse velocity near the center of the structure}
  \label{main figure label}
\end{figure}

\begin{figure}
  \centering
  \mbox{
   \subfigure[The overall scheme: $\dot{R}>0$ and $\dot{R}<0$ show the expanding and collapsing regions, respectively. \label{22kol}]{\includegraphics[width=0.5\linewidth]{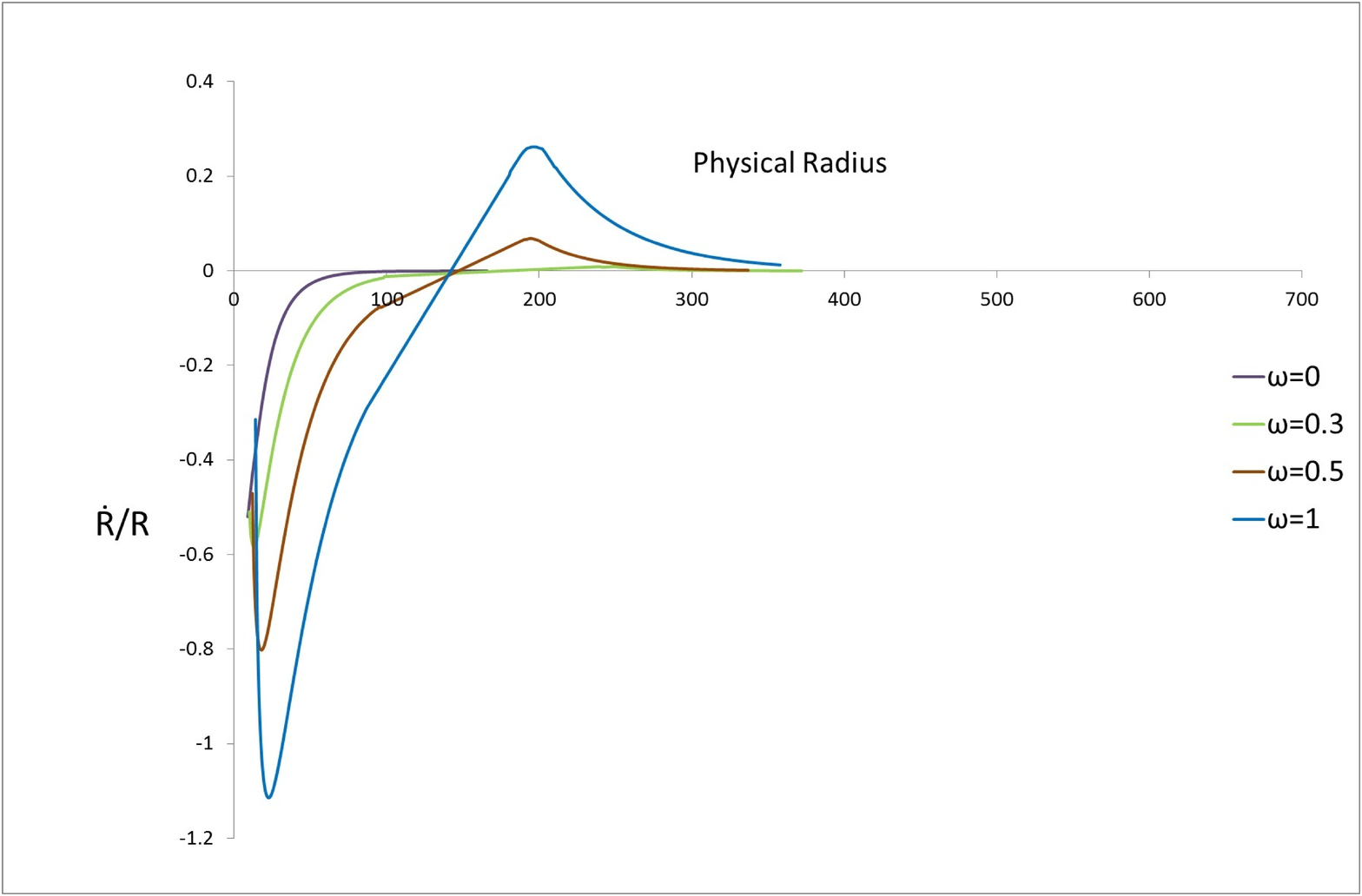}}\quad
    \subfigure[A magnified scheme related to the regions near to the center.\label{11kol}]{\includegraphics[width=0.5\linewidth]{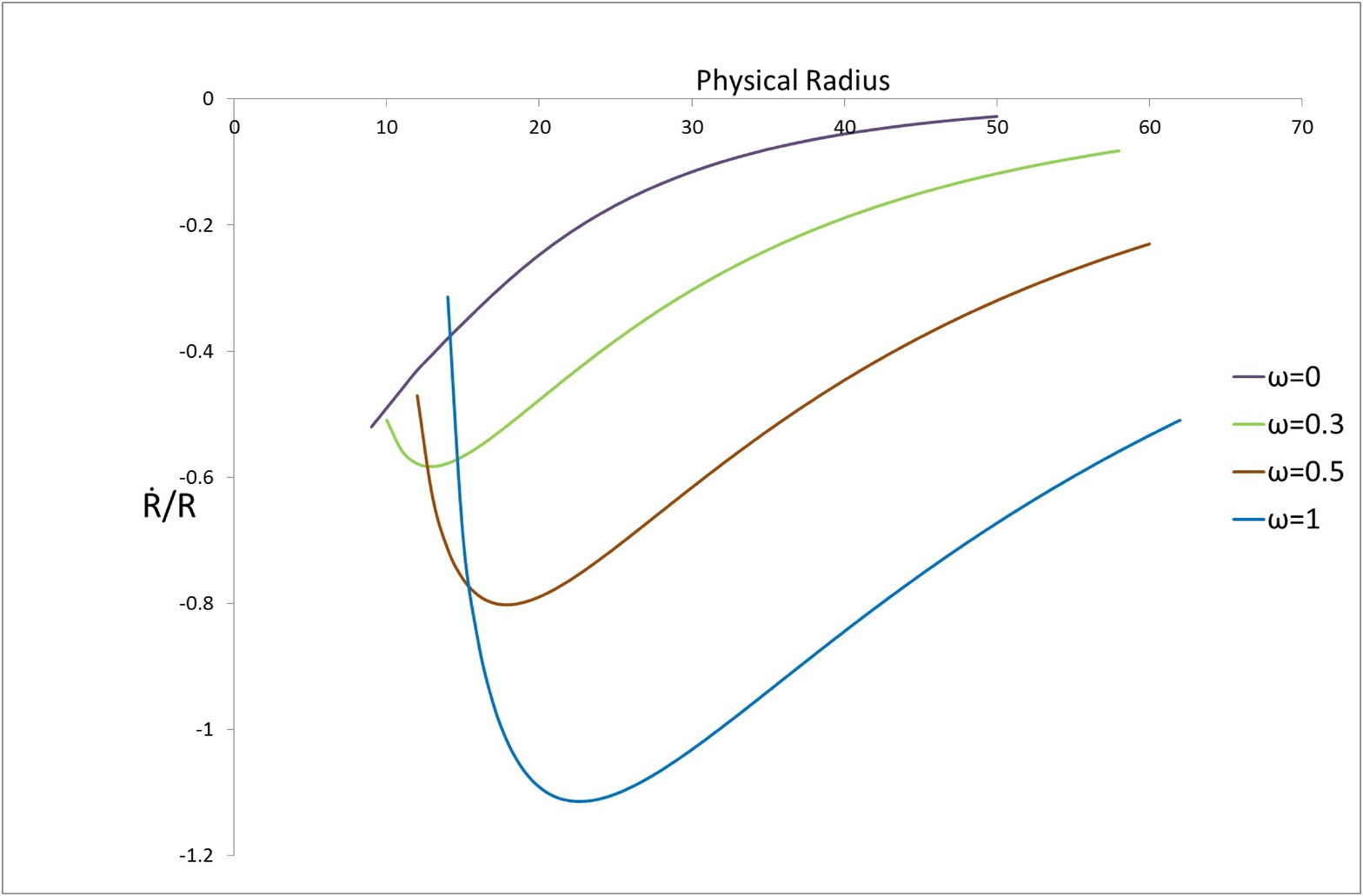}}
  }
  \caption{The behavior of the Hubble parameter $\dot{R}/R$ in the case of $p=w\rho s(r)$. The features are qualitatively as in Fig (\ref {main figure label}).}
  \label{main figure label1}
\end{figure}

\begin{figure}
  \centering
  \mbox{
   \subfigure[The overall scheme: $\dot{R}>0$ and $\dot{R}<0$ show the expanding and collapsing regions, respectively. \label{acc}]{\includegraphics[width=0.5\linewidth]{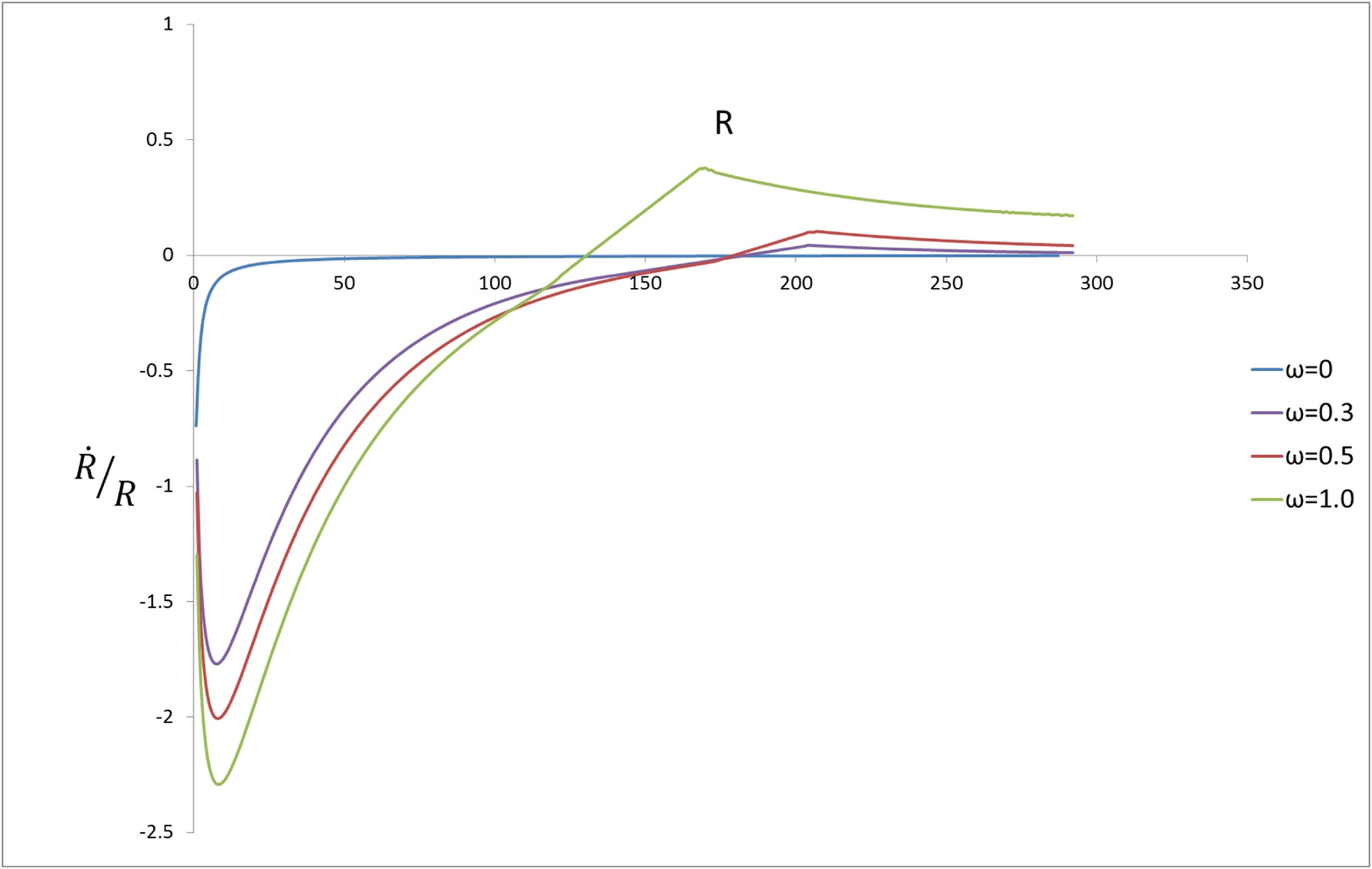}}\quad
    \subfigure[The overall scheme of the acceleration.\label{acc1}]{\includegraphics[width=0.5\linewidth]{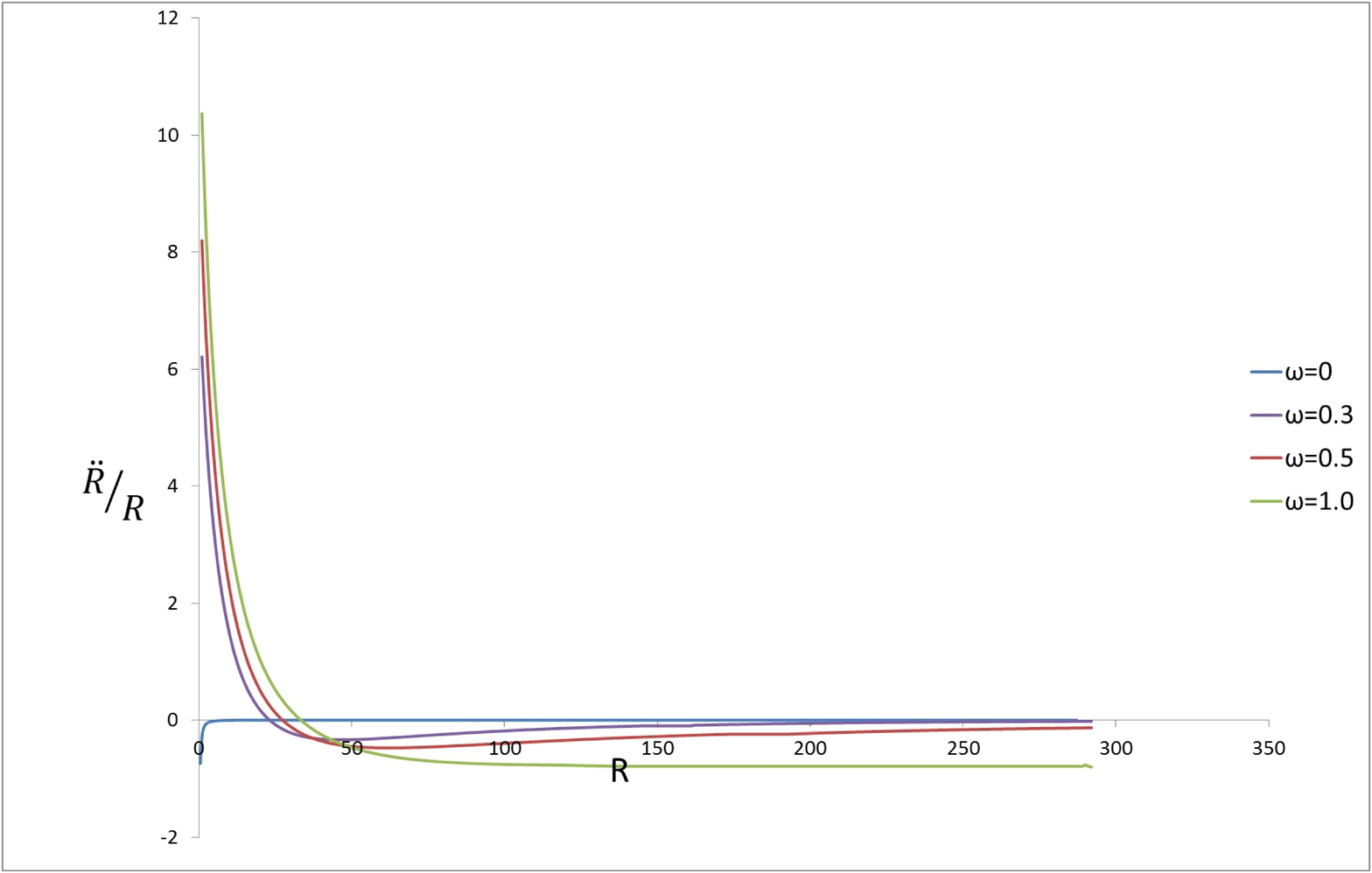}}
  }
  \caption{The behavior of the Hubble parameter $\dot{R}/R$ and acceleration $\ddot{R}/R $ in the case of $p=w\rho $. }
  \label{accelaration}
\end{figure}

\subsection{The Apparent and Event Horizon}

The boundary of a dynamical black hole, where the area law and the black hole temperature are defined, is a non-trivial concept (see for
example \cite{man} and \cite{manradiation}). Our model is again a good example to see the behavior of both apparent and event horizon of
a dynamical structure within an expanding universe. It is easily seen that the apparent horizon for our cosmological black hole is located
at $R=2M$ \cite{mangeneral}.

This apparent horizon is calculated in $t, r$ coordinates numerically.  It is always space-like tending to be light-like at late times. This can best be seen by comparing the slope of the apparent horizon relative to the light cone at every coordinate
point of it. This is in contrast to the Schwarzschild black hole horizon where it is always light-like. At the late times, however, we expect
the apparent horizon to become approximately light-like and approaching the event horizon. This is reflected in
the Figs.(\ref{horizon1}) and (\ref{horizon2}). It is evident that $ \frac{dt}{dr}|_{AH}< \frac{dt}{dr}|_{null}$ at all times on the apparent horizon, the difference tending to zero at
late times. Therefore, the apparent horizon is always a space-like \emph{dynamical horizon} leading to a \emph{slowly varying horizon} at late times \cite{ashtekar02, man}. Note
that the qualitative result is independent of the equation of state.

\begin{figure}[h]
\includegraphics[width = 7.5cm]{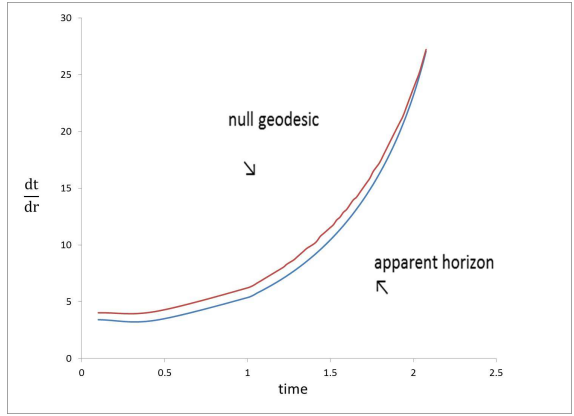}
\caption{ \label{horizon1} The $p=w\rho$  case: $ \frac{dt}{dr}|_{AH}< \frac{dt}{dr}|_{null}$  on the apparent horizon. Therefore, the
apparent horizon is always a space-like dynamical horizon leading to a slowly varying horizon at late times.}
\end{figure}

\begin{figure}[h]
\includegraphics[width = 8cm]{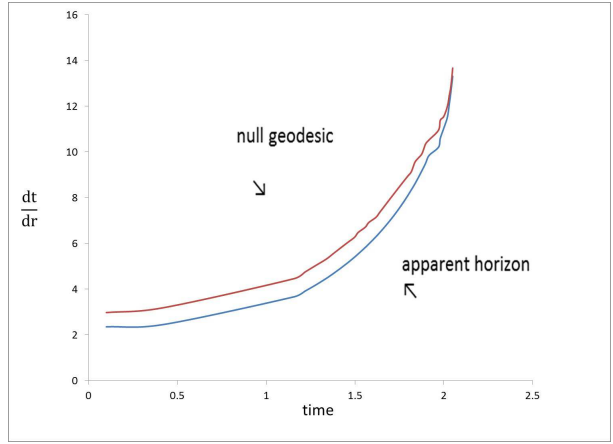}
\caption{ \label{horizon2} The $p=ws(r)\rho$ case: $ \frac{dt}{dr}|_{AH}< \frac{dt}{dr}|_{null}$  on the apparent horizon.  Qualitatively,  there
is no difference to the Fig..... }
\end{figure}

We now show how the dynamical horizon of our cosmological black hole becomes a \emph{slowly evolving horizon} at late times. Let's first define the evolution
parameter $c$ such that the tangent vector to the dynamical horizon, $V $, is given by
\begin{equation}
V^\mu=\ell^\mu-c n^\mu,
\end{equation}
where the two vectors $\ell^a$ and $n^a$ are normal null vectors on a space-like two surface $S$ in $(t,r)$ plane (see \cite{ashtekar02}).
We expect $c$ to go to zero at late times in order for our dynamical horizon to become a \emph{slowly evolving horizon}. In the case of our \L  $~$ model
$c$ is calculated to be
\begin{eqnarray}
c=2\frac{M'+w M'}{M'-w M'-R'}|_{AH}.
\label{black hole}
\end{eqnarray}
 The result of the numerical calculation  for different equation of states and different state functions is given in Figs.(\ref{slowly1}) and (\ref{slowly2}).
The decreasing behavior of the function $c$ in the course of time independent of the equation of state is evident. We may then conclude that the dynamical horizon of the
cosmological black hole tends to a slowly evolving horizon.
\begin{figure}[h]
\includegraphics[width = 8cm]{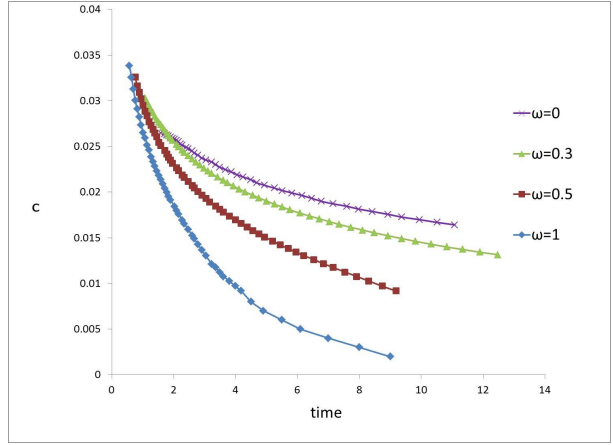}
\caption{The $p=w\rho$ case: the more pressure the sooner the dynamical horizon becomes a slowly evolving horizon. }
\label{slowly1}
\end{figure}

\begin{figure}[h]
\includegraphics[width = 8cm]{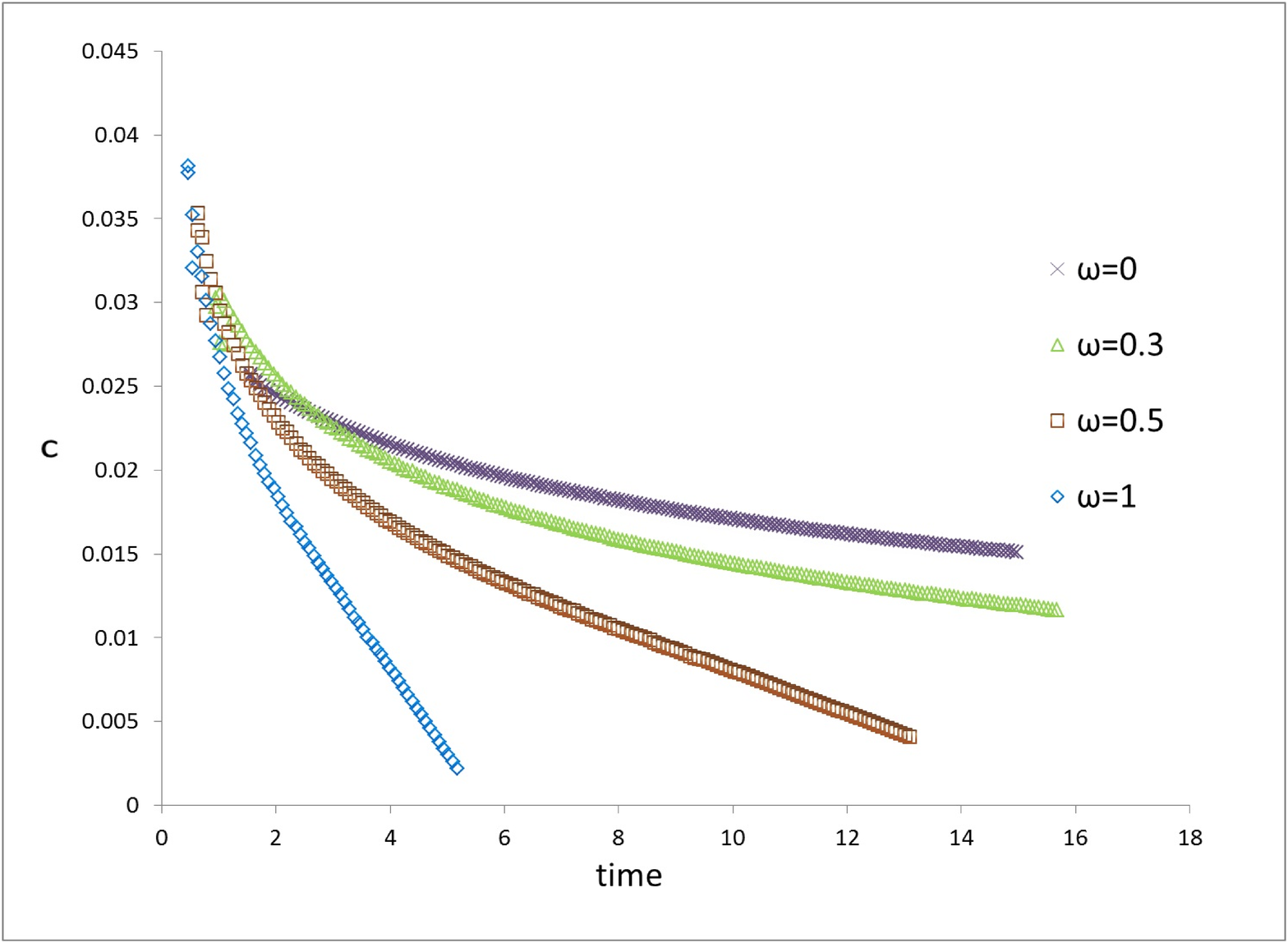}
\caption{The $p=w\rho s(r)$ case: qualitatively, the same behavior as in Fig.... }
\label{slowly2}
\end{figure}

\subsection{ Mass and matter flux}

Due to the expanding background we expect the matter flux into the dynamical black hole to be decreasing and the dynamical horizon to become a
slowly evolving horizon in the course of time\cite{mangeneral}. We know already that there is no unique concept of mass in general relativity corresponding to the
Newtonian concept. The question of what does general relativity tell us about the mass of a cosmological structure in a dynamical
setting was discussed recently \cite{manmass}. It was shown \cite{razbin} that The Misner-Sharp quasi-local mass, $M$, is very close to the
Newtonian mass.

 Let us then take the Misner-Sharp mass for this black hole and calculate the corresponding matter flux into the black
hole. In the case of \L model, the matter flux is given by
\begin{eqnarray}
\frac{dM(r,t)}{dt}|_{AH}=\frac{\partial{M(r,t)}}{  \partial{t}}|_{AH} +\frac{\partial{M(r,t)}}{\partial{r}}\frac{\partial{r}}{\partial{t}}|_{AH}=\dot{M}|_{AH}+M'\frac{\partial{r}}{\partial{t}}|_{AH}
\end{eqnarray}

 The result of the numerical calculation is depicted in Figs.(\ref{mass}) and (\ref{flux2}). Note how the pressure decreases the rate of matter flux into the
black hole.

\begin{figure}[h]
\includegraphics[width = 8cm]{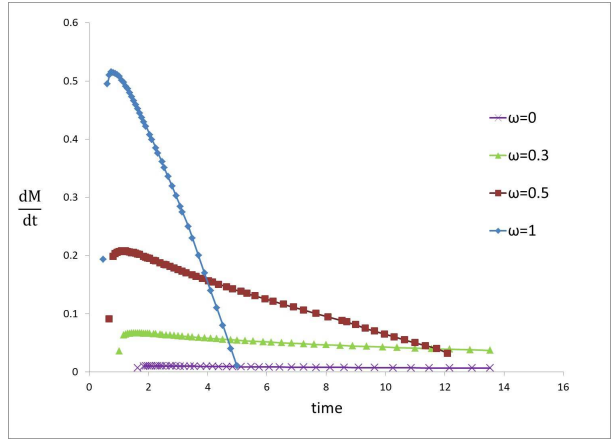}
\caption{The $p=w\rho$ case: the rate of matter flux into the black hole decreases with the pressure. }
\label{mass}
\end{figure}

\begin{figure}[h]
\includegraphics[width = 8cm]{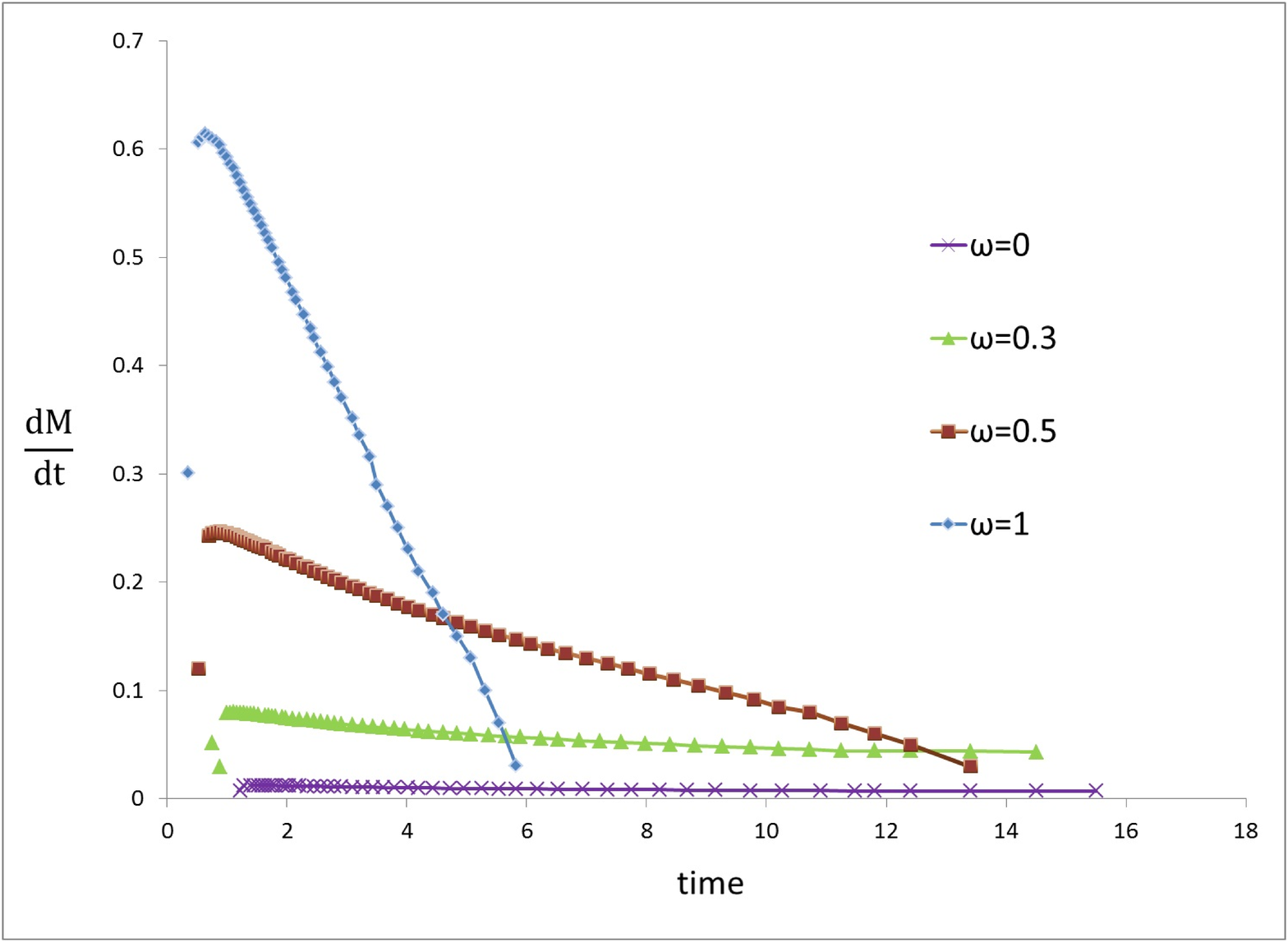}
\caption{The $p=ws(r)\rho$ case: qualitatively the same behavior as in Fig.... }
\label{flux2}
\end{figure}

\section{DISCUSSION}

We have studied the evolution of a structure made of perfect fluid with non-vanishing pressure as an exact solution of Einstein equations within an otherwise
expanding FRW universe. The structure boundary is separated by a void from the expanding part of the model which is very much like
a FRW universe already near by the void. We have noticed a counter-intuitive pressure effect somewhere inside the structure where the existence
of the pressure slows down the collapse like a classical fluid in contrast to distances far from the structure.
The collapsed region develops to a dynamical black hole with a space-like apparent horizon, in contrast to the Schwarzschild black hole. This apparent
horizon tends to a slowly evolving horizon and becoming light-like at late times with a decreasing mater flux into the black hole. We have, therefore, to
conclude that the mere existence of a cosmological matter, even dust, may have significant effect on the central black hole differentiating
it from a Schwarzschild one irrespective of how small the density outside the structure is. Hence we may not be allowed to speak about
the Newtonian approximation because of the very weak gravity in cases of non-local or quasi-local quantities such as the horizon and the mass.

\end{document}